\documentclass[amsmath,amssymb,aps,eqsecnum,nofootinbib,preprintnumbers]{revtex4}
\usepackage[dvips]{graphics,color}
\usepackage{latexsym}
\usepackage{graphicx}

\begin{document}

\preprint{KUNS-2165}
\title{Particle production in models with 
helicity-0 graviton ghost in de Sitter spacetime}

\author{
Keisuke Izumi$^a$\footnote{e-mail:
ksuke@tap.scphys.kyoto-u.ac.jp}
and 
Takahiro Tanaka$^b$\footnote{e-mail:
tanaka@yukawa.kyoto-u.ac.jp}\\~}

~\\

\address{$^a$Department of Physics, Kyoto University, Kyoto 606-8502,
Japan}
\address{$^b$Yukawa Institute for Theoretical Physics, 
Kyoto University, Kyoto 606-8502, Japan}

\begin{abstract}
We revisit the problem of the helicity-0 ghost mode of massive graviton 
in the de Sitter background. 
In general, the presence of a ghost particle, which has negative energy, 
drives the vacuum to be unstable through
pair production of ghost particles and ordinary particles. 
In the case that the vacuum state preserves 
the de Sitter invariance, 
the number density created by the pair production 
inevitably diverges due to unsuppressed ultra-violet(UV) contributions. 
In such cases one can immediately conclude that the model is not viable. 
However, in the massive gravity theory 
we cannot construct a vacuum state which respects 
the de Sitter invariance. 
Therefore the presence of a ghost does not immediately mean  
the breakdown of the model. 
Explicitly estimating the number density and the energy density of particles 
created by the pair production of two conformal scalar particles and 
one helicity-0 ghost graviton,  
we find that these densities both diverge. 
However, since models with helicity-0 ghost graviton
have no de Sitter invariant vacuum state, it is rather natural 
to consider a UV cutoff scale in the three-dimensional momentum 
space. 
Then, even if we take the cutoff scale as large as the Planck scale, 
the created number density and energy density are well suppressed.
In many models the cutoff scale is smaller than the Planck scale.  
In such models the created number density and the energy density are 
negligiblly small as long as only the physics below the cutoff scale 
is concerned. 
\end{abstract}
\maketitle

\section{introduction}

The present accelerated expansion of the Universe is 
one of the hottest topics in cosmology~\cite{SN}.
In order to explain it, 
various modified models of cosmology have been proposed and studied.
We roughly classify them into two categories.
One consists of models which utilize the spin-0 sector~\cite{darkenergy}.
Representative examples in this category are the cosmological
constant~\cite{const} and the quintessential models~\cite{quint}. 
The other category consists of models which use the spin-2 sector.
Most of modified gravity theories in this category 
fall into a massive gravity theory with higher order coupling terms.
Then, in most cases the mass of the graviton should be tuned 
to the same order as the present Hubble parameter $H_0$ if we 
try to explain the accelerated expansion of the universe.
However, it is knows that 
the helicity-0 mode of the graviton becomes a ghost mode 
in the de Sitter background with the Hubble parameter $H$
when the graviton mass is in the range $0<m^2 <2H^2$~\cite{Higuchi}.
It is often said that the existence of a ghost mode 
immediately implies that the model is not viable. 

The disaster caused by a ghost mode is easily understood 
in the Minkowski background.
The excitation energy of a ghost mode is negative~\cite{Cline,Izumi2}.
If the ghost couples with an ordinary matter field whose excitation 
energy is positive, spontaneous pair production should occur 
since it is not forbidden by the energy conservation and 
the momentum conservation.  
One may think that, if the coupling is extremely suppressed, 
the pair production rate is negligiblly small. 
However, this naive expectation is not true. 
If the initial vacuum state keeps the Lorentz symmetry unbroken, 
the probability of pair production is the same for 
the processes boosted by Lorentz transformation.
In order to calculate the total creation rate, 
we must sum up the contributions from various processed 
labelled by the 3-dimensional momentum ${\bf p}$. 
Since the integrand is $\propto 1/p$,
the integral is divergent due to UV contributions.
It will be natural to expect that in the de Sitter background 
the same pathology will remain to exist 
since the UV behavior will not be affected by the presence 
of the spacetime curvature. 

There are many attempts to construct a ghost free model in which 
a spin-2 field drives the accelerated expansion of the Universe. 
In this context DGP braneworld model~\cite{DGP} 
is the model which has been recently most extensively studied
because it has a self-acceleration branch of cosmological 
solutions~\cite{SA}.
However, the self-acceleration branch of this model is thought 
to be unrealistic, since it has a ghost mode in  
view of a four-dimensional effective theory~\cite{DGPghost}.

Nevertheless, we do not think that this is the end of the story. 
In our previous work we pointed out that 
there is no vacuum state which maintains the de Sitter invariance 
in general massive gravity theories with the helicity-0 ghost mode~\cite{Izumi2}.
Hence, the above argument which leads to  
divergent pair production in the Minkowski spacetime does not apply 
to the helicity-0 ghost mode in the de Sitter background as it is. 
In this paper, 
we examine the number density and the energy density created 
by the spontaneous production of 
two conformal scalar particles together with one helicity-0 graviton. 
The results turn out to be divergent due to the UV contributions. 
However, in massive gravity theories 
often there exists a strong coupling scale beyond which the perturbative 
expansion is no more valid.  
The strong coupling scale has been studied 
in Ref.~\cite{vainshtein} 
for generic massive gravity theories and 
in Ref.~\cite{strong} 
for DGP braneworld model with the Minkowski brane background. 
Since in the massive gravity theory 
the de Sitter invariance is already broken by 
choosing a vacuum state,   
the region of strong coupling in momentum space will be 
naturally specified not by the four-dimensional momentum 
but by the three-dimensional momentum.
In order to exclude the contribution from the region of strong coupling, 
a cutoff scale for 
the three-dimensional momentum naturally arises.
Even if we set the cutoff momentum to the Planck scale, 
the created energy density is not very large. 
The strong coupling scales estimated in literature are 
much smaller than the Planck value. 
As a result, such a model does not show violent 
particle production as long as we are restricted to the 
region where the perturbative expansion is valid. 

This paper is organized as follows.
In section~\ref{setup}, 
we will introduce a model of massive gravity theory 
with a conformal scalar field.
In section~\ref{transformation},
we discuss a transformation which 
simplifies the coupling term between the helicity-0 ghost mode and 
the scalar field. 
In section~\ref{particle}, 
we will give an estimate for the total number density and energy density of 
the created scalar particles. 
In section~\ref{disdussion},
we will summarize the results.

\section{set up}\label{setup}

We consider a massive gravity theory whose 
action is given by~\cite{FP}
\begin{eqnarray}
S =m_{pl}^2 \int d^4x \sqrt{-g} \left( R -\frac{m^2 }{4}
     (h^{\mu\nu}h_{\mu\nu} -h^2) +L_m  \right)~,
\end{eqnarray}
where $R$ is Ricci scalar, $h_{\mu\nu}\equiv g_{\mu\nu} -g_{\mu\nu}^{(0)}$ 
and $L_m$ is the matter Lagrangian.
This mass term gives the only spin-2 propagation of the graviton~\cite{FP}.
The background spacetime is given by the de Sitter metric;
\begin{eqnarray}
ds^2 &=& \left( \frac{-1}{H\eta} \right) ^2 
      (-d\eta^2 +dx^2 +dy^2 +dz^2)    \nonumber\\
   & \equiv& g^{(0)}_{\mu\nu}dx^\mu dx^\nu .
\end{eqnarray}
Since we are interested in the case with a ghost mode, 
we assume that the mass of the graviton is in the range $0<m^2<2H^2$, 
where
the helicity-0 mode of the graviton becomes a ghost mode~\cite{Higuchi}.
When we take the flat slicing of the de Sitter background, 
the helicity-0 mode of the graviton can be written as~\cite{Higuchi}
\begin{eqnarray}
&&h_{00}({\bf x}, \eta) =  \int d^3 p \frac{2p^2}{m_{pl}m \sqrt{3(2H^2-m^2)}} 
 \left( a^\dagger({\bf k})  f^{m^2}_{\bf p} (\eta) e^{i{\bf p \cdot x}} +\mbox{(h.c.)} \right),
 \label{h00} \\
&&h_{0i}({\bf x}, \eta)=
\frac{\partial_i}{\Delta}
\left[  h'_{00}({\bf x}) - \frac{2}{\eta}h_{00}({\bf x})\right]
,\\
&&h_{ij}({\bf x}, \eta) =
 \frac{\partial_i \partial_j }{\Delta^2}\biggl[\Delta h_{00}({\bf x})  
-\frac{3}{2}\left( \frac{2}{\eta} h'_{00}({\bf x}) 
+\frac{m^2 -6H^2}{(H\eta)^2}h_{00}({\bf x}) \right) \biggr] \nonumber \\
&& \qquad\qquad\qquad\qquad\qquad\qquad\qquad\qquad
+\frac{1}{2} \frac{1}{\Delta} \eta_{ij} \left[ \frac{2}{\eta}  h'_{00}({\bf x}) 
+ \frac{m^2 -6H^2}{(H\eta)^2}h_{00}({\bf x}) \right],
\end{eqnarray}
with
\begin{equation}
f^{m^2}_{\bf k} ({\bf x},\eta) 
= \left( \frac{\pi H^2}{4}e^{-\pi Im (\nu)} \right) ^{1/2}
          \eta^{\frac{3}{2}} H^{(2)}_\nu ( k \eta)~,
\label{f}
\end{equation}
where a prime ``$~'~$'' denotes a differentiation with respect to
$\eta$. 
The metric component $h_{00}$ satisfies the equation of the Klein-Gordon type 
\begin{eqnarray}
\left(-\partial_\eta^2+\partial_x^2+\partial_y^2+\partial_z^2
+\frac{2}{\eta}\partial_\eta -\frac{m^2}{(H\eta)^2}\right) h_{00}=0. 
\label{hEOM}
\end{eqnarray}
Notice that, in order to keep the commutation relation 
$[a({\bf k}),a^\dagger({\bf k'})]= \delta^3({\bf k}-{\bf k'})$, 
it is necessary to associate the positive (negative) frequency mode functions 
in the normal sense with the creation (annihilation) operators
$a^\dagger({\bf k'})$ ($a({\bf k})$)~\cite{Izumi2}.  

For simplicity, we consider a conformal scalar field 
as a normal non-ghost matter field. 
The action is given by 
\begin{eqnarray}
S_{m} =\int d^4x \frac{1}{2} \sqrt{-g} \left\{ g^{\mu \nu} 
\partial _\mu \phi \partial_\nu \phi - \frac{1}{6} R  \phi^2 \right\} ~, 
\label{Sconf}
\end{eqnarray}
which inherently has the coupling to the graviton. 
Following the canonical quantization,
we can write $\phi$~\cite{Higuchi}
\begin{eqnarray}
&&\phi = \int  d^3 k \left( b({\bf k}) \Phi({\bf k}) e^{i {\bf k\cdot
		      x}}+ \mbox{(h.c.)} \right),
\end{eqnarray}
where
\begin{equation}
\Phi({\bf k})= f^{2H^2}_{\bf k} ({\bf x},\eta)
=\frac{H}{\sqrt{2k}} \eta \exp (-ik \eta), \label{Phi}
\end{equation}
and the creation and annihilation operators satisfy
\begin{eqnarray}
&&[b({\bf k}),b^\dagger({\bf k'})] = \delta^3({\bf k}-{\bf k'}),\\
&&[b({\bf k}),b({\bf k'})] =[b^\dagger({\bf k}),b^\dagger({\bf k'})] =0.
\end{eqnarray}

The leading order coupling between 
the conformal scalar field and the graviton
can be deduced from Eq.(\ref{Sconf}) as
\begin{eqnarray}
S_{int} = -\frac{1}{2}\int d^4 x \sqrt{-g^{(0)}} h^{\mu \nu} T_{\mu \nu}, 
\label{Sint}
\end{eqnarray}
where
\begin{eqnarray}
&&T_{\mu \nu} =\frac{2}{3}\phi_{;\mu } \phi_{;\nu } 
  -\frac{1}{6} g_{\mu \nu } g^{\rho \sigma} \phi_{;\rho} \phi_{;\sigma}
 -\frac{1}{3}\phi_{;\mu \nu }\phi 
  +\frac{1}{12} g_{\mu \nu} \phi \Box \phi,
\end{eqnarray}
is the energy-momentum tensor of 
the conformal scalar field.

\section{transformation of coupling term}\label{transformation}


The mode function of the helicity-0 mode of the graviton
with the three dimensional momentum ${\bf p}=(p,0,0)$
can be written as
\begin{eqnarray}
h_{\mu \nu} ({\bf p})=  \left(
  \begin{array}{cccc}
    f_1   & ipf_2   & 0   & 0   \\
    ipf_2   &  -p^2f_3 +f_4  & 0   & 0   \\
     0  &  0  & f_4 & 0   \\
     0  &   0 &  0  &  f_4  \\
  \end{array}
\right)  a^\dagger ({\bf k}) \exp ( ipx) +\mbox{(h.c.)} , 
\nonumber\\
&&\qquad
\end{eqnarray}
where 
\begin{eqnarray}
&&f_1(p,\eta)= \frac{2p^2}{m_{pl}m \sqrt{3(2H^2-m^2)}}  f^{m^2}_{\bf p}  (\eta),
\\
&&f_2(p,\eta)=-\frac{1}{p^2}
\left[  f'_1(p,\eta) - \frac{2}{\eta}f_1(p,\eta)\right],
\\
&&f_3(p,\eta)=-\frac{1}{p^4}\biggl[p^2 f_1(p,\eta)  
+\frac{3}{2}\left( \frac{2}{\eta}  f'_1(p,\eta) 
+\frac{m^2 -6H^2}{(H\eta)^2}f_1(p,\eta) \right) \biggr] ,
\\
&&f_4(p,\eta)=-\frac{1}{2 p^2} \left[ \frac{2}{\eta}  f'_1(p,\eta) 
+ \frac{m^2 -6H^2}{(H\eta)^2}f_1(p,\eta) \right] .
\end{eqnarray}
We consider the transformation defined by 
\begin{eqnarray}
\hat h_{\mu \nu} = h_{\mu \nu} +\nabla _\mu \xi_\nu +\nabla _\nu \xi_\mu . 
\label{trans}
\end{eqnarray}
Because the massive gravity theory has no gauge degree of freedom,
this transformation changes the form of the action.
However, using the conservation law $\nabla ^\mu T_{\mu \nu}=0$,
one can show 
\begin{eqnarray}
\int d^4 x \sqrt{-g} ( \nabla ^\mu \xi^\nu ) T_{\mu \nu} 
=\int d^4 x \sqrt{-g} \xi^\nu  \nabla ^\mu T_{\mu \nu} \nonumber 
=0 .
\end{eqnarray}
This means that the shape of the leading order interaction terms does not change 
under this transformation. 
Using this relation, we simplify the lowest order coupling 
(\ref{Sint})\footnote{
The change caused by this transformation is to add total derivative 
terms with respect to the time coordinate $\eta$.
This change corresponds to the canonical transformation of 
variables. }.
The helicity-0 component of $\xi$ can be written as
\begin{eqnarray}
\xi_\mu = 
\left(    A   , ipB   ,  0  ,  0  
\right)
 \exp (-ipx) .
\end{eqnarray}
By setting 
\begin{eqnarray}
A = -f_2-\frac{1}{2}  f'_3-\frac{1}{\eta} f_3 \qquad
\mbox{and} \qquad
B = -\frac{f_3}{2}  ,
\end{eqnarray}
we obtain 
\begin{eqnarray}
\hat h_{\mu \nu} 
= diag \Bigl( f_1 +\frac{2}{\eta}A +2A' , f_4+\frac{2}{\eta}A 
, f_4+\frac{2}{\eta}A, f_4+\frac{2}{\eta}A \Bigr) +\mbox{(h.c.)}.
\label{eq:c3}
\end{eqnarray}
We decompose $\hat h_{\mu\nu}$ into the pure trace component $h^T_{\mu \nu}$ and 
the $\{00\}$-component $h^S_{\mu \nu}$ as
\begin{eqnarray}
&&\hat h _{\mu \nu} = h^T_{\mu \nu} +h^S_{\mu \nu} , \label{hath} \\
&&h^T_{\mu \nu} =diag \left( -f_4-\frac{2}{\eta}A,f_4+\frac{2}{\eta}A,
f_4 +\frac{2}{\eta}A,f_4+\frac{2}{\eta}A \right)+\mbox{(h.c.)} , 
\\
&&h^S_{\mu \nu} = diag \left( f_1+f_4+\frac{4}{\eta}A +2A',0,0,0 \right)+\mbox{(h.c.)}.
\end{eqnarray}
Then, as the energy-momentum tensor of the conformal scalar field 
is traceless, the interaction term becomes
\begin{eqnarray}
&&S_{int} 
=-\frac{1}{2} \int d^4 x h^S_{00} T_{00}.
\end{eqnarray}
and from Eqs.~(\ref{h00})-(\ref{f}) we have 
\begin{eqnarray}
&& h^S_{00} = \int d^3 p\frac{m \sqrt{3(2H^2 -m^2 )}}
 {m_{pl} H^4 \eta^4 p^2}
 \left( a^\dagger({\bf k})  f^{m^2}_{\bf k} (x,\eta) + h.c \right),
\end{eqnarray}
where we used the on-shell condition (\ref{hEOM})%
\footnote{
As we consider only the lowest order effect in coupling 
in this paper, 
Feynman diagrams containing internal loops are neglected. 
Therefore the interaction term can be rewritten by using 
the on-shell condition~(\ref{hEOM}).
}.

\section{particle creation from the vacuum}\label{particle}

\begin{figure}[tbp]
  \begin{center}
    \includegraphics[keepaspectratio=true,height=40mm]{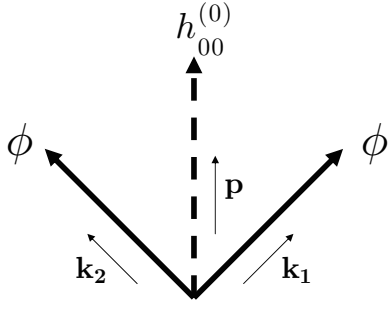}
  \end{center}
  \caption{a diagram of the lowest order pair creation of
            $h^{(0)}$ and $\phi$.}
  \label{fig:coup.eps}
\end{figure}

In this section, 
we estimate the number density of $\phi$-particles 
created through the process lowest order in coupling,  
which is diagrammatically expressed in Fig.~\ref{fig:coup.eps}.
The total number of the created  $\phi$-particles 
with the momentum ${\bf k}$ will be evaluated by 
taking the expectation value of 
the number operator $N_{\phi,{\bf k}}\equiv b^\dagger({\bf
k})b({\bf k})$, which is calculated at the leading order of perturbation
as%
\footnote{
Generally speaking, in the canonical quantization, 
the interaction action $S_{int}$ is not identical to the spacetime 
integral of the non-linear term of the Lagrangian 
if the kinetic terms are not canonical. 
However, in the computation at the leading order of perturbation 
the difference does not arise~\cite{Seery}.
}
\begin{eqnarray}
&&_{in} \langle\, 0 \,|\,  N_{\phi,{\bf k}} \,|\,  0 \,\rangle_{in}
=\int d^3\! k_1\,  d^3\! k_2 \, d^3\! p_1 \, d^3\! k_3 \, d^3\! k_4 \, d^3\! p_2\, 
\langle\, 0 \,|\,  S_{int} \,|\,  {\bf k}_1 {\bf k}_2 {\bf p}_1 \,\rangle \cr
&&\qquad \qquad\qquad\qquad\qquad\qquad\qquad\qquad
\times \langle\, {\bf k}_1 {\bf k}_2 {\bf p}_1 \,|\,  N_{\bf k} 
\,|\,  {\bf k}_3 {\bf k}_4 {\bf p}_2 \,\rangle
\langle\, {\bf k}_3 {\bf k}_4 {\bf p}_2 \,|\,  S_{int}\,|\,  0\,\rangle,\label{Nphi}
\end{eqnarray}
where $\{k_i\}$ and $\{p_i\}$ are the momenta of 
$\phi$-particles and ghost particles, respectively.
Using the relations 
\begin{eqnarray}
&&
{\Phi'} ^*=\frac{1}{\eta}\Phi^* +ik \Phi^*,
\\
&&
{\Phi''}^*=ik\frac{2}{\eta}\Phi^* -k^2 \Phi^*,
\end{eqnarray}
derived from Eq.~(\ref{Phi}), 
and the momentum conservation law
\begin{eqnarray}
{\bf k_1}\cdot{\bf k_2}=\frac{p^2-k_1^2-k_2^2}{2},
\end{eqnarray}
we can evaluate the matrix element as
\begin{eqnarray}
&&\langle\, {\bf k}_1, {\bf k}_2, {\bf p}_1 \,|\,  S_{int}\,|\,  0\,\rangle\nonumber \cr
&&\qquad
= \langle\,{\bf k}_1, {\bf k}_2, {\bf p}_1 \,|\,  \int_{-\infty}^{\eta_f}  
d \eta\,  d^3 \! k_3\, d^3 \! k_4\, d^3 \! p_2 \  
\lambda ({\bf k}_3,{\bf k}_4,{\bf p}_2) \delta^{(3)}\! 
\left({\bf k}_3 +{\bf k}_4 - {\bf p}_2\right)  
b^\dagger ({\bf k }_3) b^\dagger({\bf k}_4) a^\dagger ({\bf p}_2)
\,|\,  0\,\rangle\nonumber \cr
&&\qquad
= 2 \int_{-\infty}^{\eta_f} d \eta  \ \lambda ({\bf k}_1,{\bf k}_2,{\bf p_1}) 
\delta^{(3)}i\! \left({\bf k}_1 +{\bf k}_2 - {\bf p_1}\right) ,\label{Sint}
\end{eqnarray}
where
\begin{eqnarray}
\lambda({\bf k_1},{\bf k_2},{\bf p})=-\frac{m\sqrt{2H^2-m^2}\bigl(3(k_1-k_2)^2-p^2\bigr)}
{16\sqrt{3}m_{pl}H^2\sqrt{k_1 k_2}p^2 \eta^2}
 f^{m^2}_{\bf p}e^{i(k_1+k_2)\eta} .
\end{eqnarray}
On the other hand, 
$\langle\,{\bf k}_1, {\bf k}_2, {\bf p}_1 \,|\,  N_{\bf k} \,|\,  {\bf k}_3, {\bf k}_4, {\bf p}_2 \,\rangle $
is given by 
\begin{eqnarray}
&&\langle\,{\bf k}_1, {\bf k}_2, {\bf p}_1 \,|\,  N_{\bf k} \,|\,  {\bf k}_3, {\bf k}_4, {\bf p}_2\,\rangle 
= \langle\, 0 \,|\,  a({\bf p}_1) b({\bf k}_2) b({\bf k}_1) b^\dagger ({\bf k}) b({\bf k}) 
 b^\dagger ({\bf k}_3) b^\dagger ({\bf k}_4) a^\dagger ({\bf p}_2) \,|\,  0\,\rangle\nonumber \\
&&\qquad \qquad\qquad\qquad\qquad\quad\ \ \!
 = \delta^{(3)}\!({\bf p}_1 -{\bf p}_2) 
    \left\{ \delta^{(3)}\!({\bf k}_1 -{\bf k}) +\delta^{(3)}\!({\bf k}_2 -{\bf k}) \right\} \nonumber \\
&&\qquad \qquad \qquad\qquad\qquad\qquad\qquad
 \times \left\{ \delta^{(3)}\!({\bf k}_1 -{\bf k}_3) \delta^{(3)}\!({\bf k}_2 -{\bf k}_4) +
                  \delta^{(3)}\!({\bf k}_1 -{\bf k}_4) \delta^{(3)}\!({\bf k}_2 -{\bf k}_3)\right\}.
\label{N} 
\end{eqnarray}
Combining Eqs.~(\ref{Nphi}), (\ref{Sint}) and (\ref{N}),
we find that the number of $\phi$-particles per unit comoving volume  
is given by 
\begin{eqnarray}
&&n_{com} \equiv 
\int d^3\!k\, \frac{_{in} \langle\, 0 \,|\,  N_{\bf k} \,|\,  0 \,\rangle_{in}}
{ \delta^{(3)}\!(0)}
\nonumber \\
&&\qquad\ 
= 16 \int d^3\!k\, d^3\! k_1\, d^3\! p\, \left| \int_{-\infty}^{\eta_f} d\eta \ 
\lambda ({\bf k},{\bf k}_1,{\bf p}) \right| ^2  \delta^{(3)}\! 
\left({\bf k} +{\bf k}_1 - {\bf p} \right) .\label{nd}
\end{eqnarray}

Since the pathology caused by the existence of a ghost mode 
is that the number or energy density of the created particles 
suffers from UV divergence, 
we concentrate on the behavior in the UV limit.
In this limit,  
we can use the approximation
\begin{eqnarray}
f^{m^2}_{\bf p} \simeq  \frac{H \eta}{\sqrt{2 p}}
e^{-ip\eta}.
\end{eqnarray}
Then, we have 
\begin{eqnarray}
\int_{-\infty}^{\eta_f} d\eta \ \lambda ({\bf k},{\bf k}_1,{\bf p}) \simeq
-\frac{m\sqrt{2H^2-m^2}\bigl(3(k-k_1)^2-p^2\bigr)}
{16\sqrt{6}m_{pl}H\sqrt{k k_1 p^5}}\int_{-\infty}^{\eta_f} d\eta 
\frac{1}{\eta} e^{i(k+k_1-p)\eta}. \label{lambda}
\end{eqnarray}

Since the background is not stationary, 
the energy conservation law does not hold.
Hence, Eq.~(\ref{lambda}) contains 
the contributions not only from the ghost instability 
but also from the violation of the energy conservation law. 
The latter contribution exists even if there is no ghost
excitation~\cite{Higuchi2} and is divergent. 
However, we think that the divergence of this type is responsible for 
the uncertainty in the definition of a particle when the 
interaction is turned on. In fact, if we smoothly turn off 
the interaction before $\eta_f$, UV divergence disappears 
in the non-ghost case. 
In contrast, 
the UV contribution due to the presence of a ghost mode does not disappear 
even if we smoothly turn off the interaction. 
Such a contribution comes from the momentum 
region $(k+k_1-p)|\eta_f| \ll 1$. In this region, roughly speaking, 
the integral in Eq.~(\ref{lambda}) is $O(1)$. 
Then, we can easily estimate Eq.~(\ref{lambda}) as
\begin{eqnarray}
\int_{-\infty}^{\eta_f} d\eta \ \lambda ({\bf k},{\bf k}_1,{\bf p}) 
\biggr|_{\mbox{\scriptsize ghost contribution}} \cong
\frac{m\sqrt{2H^2-m^2}\bigl(3(k-k_1)^2-(k+k_1)^2\bigr)}
{m_{pl}H\sqrt{k k_1 (k+k_1)^5}}~.
\label{lam}
\end{eqnarray}
Substituting this estimate into Eq.~(\ref{nd}), 
we have
\begin{eqnarray}
n_{com} 
&\cong & 
\int dk\, dk_1\, d\!\cos\! \theta\, k^2 k_1^2 
 \frac{m^2(2H^2-m^2)\bigl(k^2-4 k k_1+k_1^2\bigr)^2}
{m_{pl}^2 H^2 k k_1 (k+k_1)^5}   ,\label{gc}
\end{eqnarray}
The range of the $\cos \theta$-integral in the above expression 
is approximately given by $(k+k_1-|{\bf k}+{\bf k}_1|)|\eta_f| \le 1$, 
which leads to 
$1- {k+k_1\over k k_1 |\eta_f|}\le \cos \theta \le 1$.
Then the number density is estimated as 
\begin{eqnarray}
n_{com} = O\left( {m^2(2H^2-m^2) \Lambda^2 \over m_{pl}^2 H^2 |\eta_f|} \right) ,
\end{eqnarray}
where we have introduced a cutoff $\Lambda$ in the 
three-dimensional comoving momentum integral. This result means that 
the number density is quadratically divergent.
In the same manner the energy density of the created $\phi$-particles is
also evaluated as%
\begin{eqnarray}
\rho_{com} \cong 
\int d^3k \frac{_{in} \langle\, 0 \,|\,  N_{\bf k}\,  k \,|\, 0 \,\rangle_{in}}
{ \delta^{(3)}\!(0)} 
= O\left( {m^2(2H^2-m^2) \Lambda^3 \over m_{pl}^2 H^2
	       |\eta_f|} \right).
\end{eqnarray}

\section{discussion}\label{disdussion}

We have studied the helicity-0 ghost of massive graviton 
in de Sitter space time.
It is often said that the existence of a ghost mode 
immediately means that the model is not viable.  
In the Minkowski background this is because 
infinitely many particles are instantaneously created 
through the pair production of a ghost and a normal particle, 
irrespective of the strength of   
the interaction between the ghost and the normal particle.  
In de Sitter background the same phenomena is expected to occur
because the UV behavior is almost the same.
However, since the massive gravity theory with a ghost mode
has no vacuum state which respects the de Sitter invariance, 
the argument used in the case of Minkowski background, which assumes 
the Lorentz invariance of the initial vacuum state, does not apply. 
Therefore, in this paper, 
we have explicitly evaluated the number density and the energy density of the
particles created by the pair production, 
taking a conformal scalar field as the matter content which 
couples to the helicity-0 mode of the graviton. 

The result was divergent due to the UV contribution.
However, in the modified gravity theory, there is a natural cutoff
momentum scale 
beyond which the model is strongly coupled. 
Since non-perturbative effects become important beyond the strong coupling scale, 
the region in the momentum space where the linear theory is justified is restricted.
Since the vacuum state does not have de Sitter invariance, 
it is not so strange even if the model has a three-dimensional 
momentum cutoff instead of the usual four-dimensional covariant one. 
If the three-dimensional momentum cutoff is set to the Planck scale, 
which means $\Lambda H \eta_f=m_{pl}$,  
the proper energy density of the created particles becomes 
\begin{eqnarray}
\rho= \rho_{com} H^4 \eta_f^4 
\alt O\left( H^3 m_{pl}   \right), 
\end{eqnarray}
where we used the fact that both $m^2$ and $(2H^2-m^2)$ are $\alt O(H^2)$. 
This energy density is much smaller than the critical energy density of the universe 
$\rho_{crit} = H^2 m_{pl}^2$.
Since we expect that the three-dimensional momentum cutoff scale 
is generally smaller than the Planck scale, 
the particle creation is extremely suppressed 
unless we extrapolate the result of perturbative analysis beyond 
its validity region.
In order to discuss the possible hazardous nature of the helicity-0
ghost in the large momentum region, 
we need a method to handle non-perturbative quantum effects.  
We think without such a method one cannot conclude that
models with a helicity-0 ghost in the massive gravity theory 
are all to be excluded. 

In this paper, for simplicity, we have considered only a conformal scalar field.
However, there could be a possibility 
that the UV behavior is worse for non-conformal fields 
because  the conformal invariance eliminates the coupling through 
the trace of the energy momentum tensor. 
We will discuss the cases of more generic matter contents 
in future.

\acknowledgements
The authors thank Takashi Nakamura for his valuable comments and continuous
encouragement.
TT is supported by Grant-in-Aid for
Scientific
Research, Nos. 19540285 
and by Monbukagakusho Grant-in-Aid
for Scientific Research(B) No.~17340075. 
We also acknowledge the support from
the Grant-in-Aid for the Global COE Program gThe
Next Generation of Physics, Spun from Universality and
Emergenceh from the Ministry of Education, Culture,
Sports, Science and Technology (MEXT) of Japan.\\

\end{document}